# WIND SPEED PREDICTION BY DIFFERENT COMPUTING TECHNIQUES

## Munir Ahmad Nayak[1]  and  M C Deo[2]

[1] M Tech candidate, Indian Institute of Technology Bombay, Powai, Mumbai 400 076
([munir.nit70@gmail.com](munir.nit70@gmail.com))
[2] Professor, Indian Institute of Technology Bombay, Powai, Mumbai 400 076
([mcdeo@iitb.ac.in](mcdeo@iitb.ac.in))

**ABSTRACT:**

Wind is slated to become one of the most sought after source of energy in future. Both onshore as well as offshore wind farms are getting deployed rapidly over the world. This paper evaluates a neural network based time series approach to predict wind speed in real time over shorter durations of up to 12 hr based on analysis of three hourly wind data collected through a wave rider buoy deployed off Goa in deep water and far away from the shore. The data were collected for 4 years from February 1998 to February 2002.

A simple feed forward type of network trained using a variety of algorithms was used. The input nodes selected by trial were three in number and belonged to the segment of preceding observations while the output node was single and it consisted of the predicted value of the wind speed over the subsequent 3, 6 and 12 hours – one at a time. The number of hidden nodes was based on trials. The total sample was divided into a training set (first 70 percent) and a testing set (remaining 30 percent). The outcome of the network was compared with the actual observations with the help of scatter diagrams and time history plots as well as through the error statistics of the correlation coefficient, R, and mean square error, MSE. The testing of the network showed that it predicted the wind speed in a very satisfactory manner with R = 0.99 and MSE = 0.30 $(m/s)^2$ for a 3-hour ahead prediction while these values for a 12-hour ahead predictions were 0.96 and 1.19 $(m/s)^2$, respectively. Such a prediction based on neural network was found to be superior to that based on polynomial fittings as well as ARMA models. ARIMA models were also used but the predicted values showed significant lag.

**Introduction**

Wind is the mass movement of air due to the difference in pressure between two sections on earth. Wind is characterized by its speed, direction, time of occurrence mainly. Wind force is given by a scale called the "Beaufort wind force scale" which provides an empirical description of wind speed based on observed sea conditions. Wind produces waves in the ocean which have a lot of significance in real world.

Wind energy is infinite and inexhaustible and its use in energy production does not lead to any pollution and hence is a better way to produce energy without being against environment. Better techniques need to be adopted for efficient usage of wind energy. The most important factor which influences wind energy production is the local wind speed[1]; for which there is a great need of development of improved forecasting methods which will directly improve the transmission of energy and resource allocation and hence will determine the reliability of energy producing company and the operation of energy production systems and energy distribution[2]. Not only in case of energy production but wind plays an important role in military and civilian fields for air traffic control. Real-time prediction of wind over duration of next few hours has applications in aircraft operations, wind farm monitoring and issuing storm warnings to citizen.

**Data used and station location**

The data used for case study is the wind speed at station Goa shore. Data is available from February 1998 to February 2002. The data is available in time intervals of three hours. Eventually three hour interval was taken for case study. The station is having latitude of 15.447N and longitude of 69.236E. The average water depth in the sea is 3800m and distance from shore of the measurement station is 492km.

The modeling techniques which are used are; polynomial curve fitting, ARMA, and artificial neural networks. ARIMA models were also used but they were giving unexpected results and hence were not considered for further use. These are discussed one by one and case study for the above station is done by all the three models developed.

**Polynomial curve fitting**

Curve fitting technique is used in many application of data analysis. It is advantageous if the data is following some pattern with less randomness about the mean. Curve fitting is actually developing an implicit equation for determining one variable with respect one or more variables [3]. Polynomial curve fitting can be considered as Generalized Linear Least-Squares Regression [4].

**The Method Mathematically**

A general implicit equation for a variable 'θ' in terms of variables x and y where $(x_i, y_i)$ is experimental data i=1, 2…n. can be written as follows:
$$P(x; y; \theta) = 0 \quad (1)$$
Here P refers to polynomial. The coefficients of polynomial need to be determined for fitting the curve. The degree of the curve will determine the number of coefficients. In the present study the independent variable is the past data only. Accordingly our polynomial curve will be having only one parameter for estimating 'θ'. In simple statement
$$\Theta = f(y) \quad (2)$$
Θ is the wind speed m/s corresponding to any time 't', and y is the previous wind speed. In simpler way
$$\Theta = a_0 + a_1 y + a_2 y^2 + a_3 y^3 + \ldots\ldots + a_n y^n \quad (3)$$
Using the available 70% data; polynomial equations were found with different degrees and corresponding mean square error and correlation coefficients were determined.
If the observed valves of θ are $\theta_i$; where i= 1, 2 ………n
For the best fit the residual should be least and is called the least square. The polynomial curve fitting is a generalized form of least square method.

Coefficient of correlation is given as:
$$R = \sum[(\theta-\theta')(f(y)-f'(y))] / \sqrt{[\sum\{(\theta-\theta')^2\}\sum\{f(y)-f'(y)\}^2]} \quad (4)$$

Mean square error is given as
MSE = $\sum [\theta_i - f(y)]^2 / n$

(5)

**Case study**

The data was analyzed and several model polynomial curves are fitted and out of those two and three degree curves are found to fit best. The equations of curves given below are the best fit polynomial curves for the data available.

**Case study models**

$$\theta_{t+p} = a_n \theta_t^n + a_{n-1} \theta_t^{n-1} + \ldots + a_0$$

| Time ahead | Best model | Coefficients | Mean square error (MSE) | Correlation coefficient (r) |
|---|---|---|---|---|
| 3 hour ahead | $2^{nd}$ degree | $a_2$ : 0.0045<br>$a_1$ : 0.8930<br>$a_0$ : 0.7173 | 1.4243 | 0.9182 |
| 6 hour ahead | $3^{rd}$ degree | $a_3$ : -0.0018<br>$a_2$ : 0.05550<br>$a_1$ : 0.38150<br>$a_0$ : 1.85420 | 1.7781 | 0.8788 |
| 12 hour ahead | $3^{rd}$ degree | $a_3$ : -0.0024<br>$a_2$ : 0.07220<br>$a_1$ : 0.18680<br>$a_0$ : 2.48350 | 2.5341 | 0.8221 |

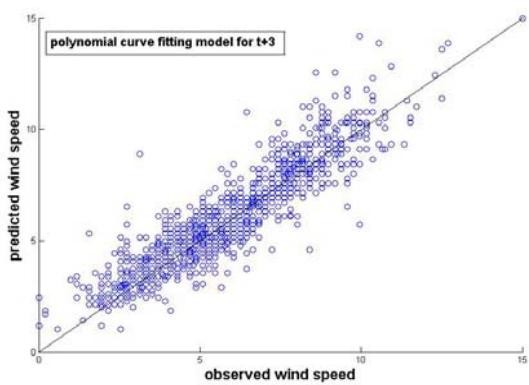
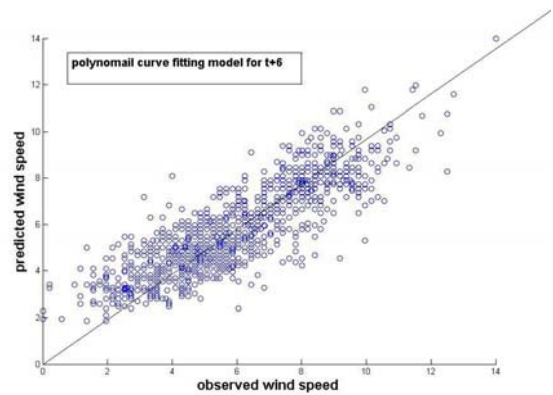

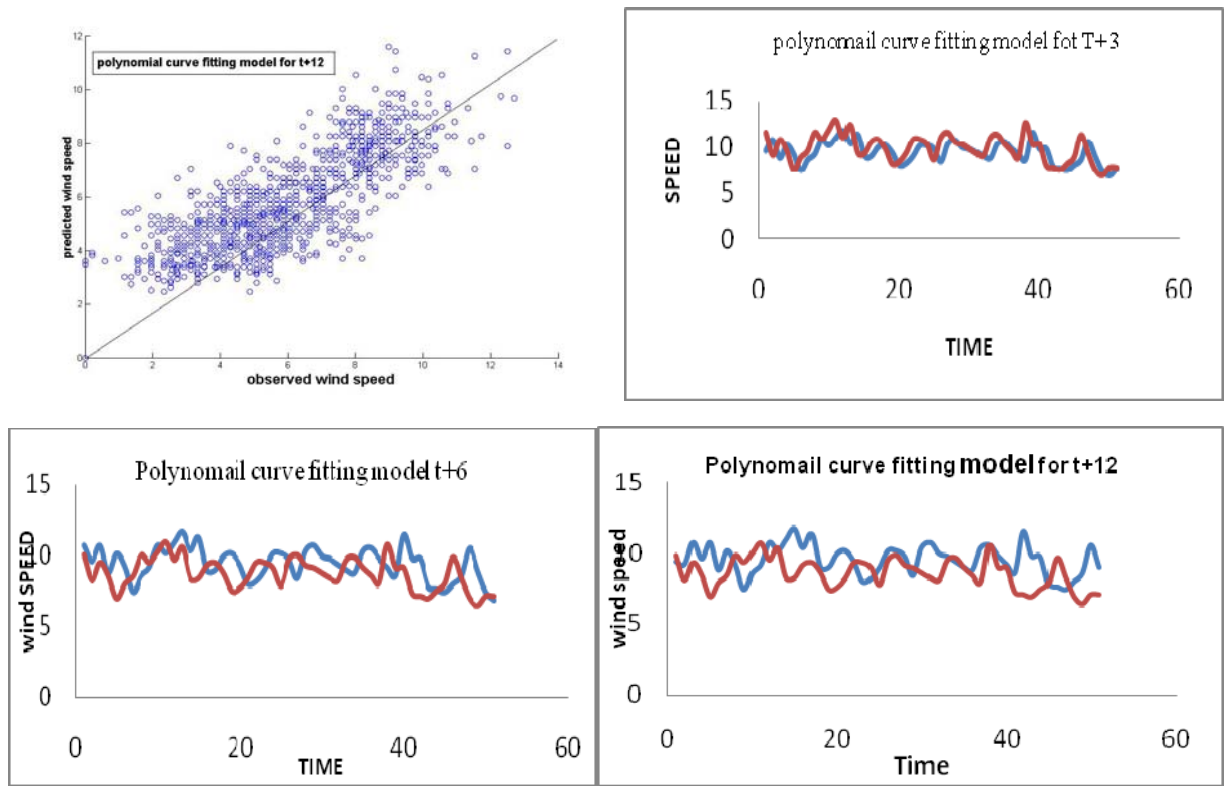

**Figure 1** Comparison of observed wind speed and predicted wind speed for 3, 6 and 12 hour ahead wind speed prediction with polynomial curve fitting models from "01/08/2001 3:15" to "16/12/2001 12:15".

**Auto-Regressive Moving Average (ARMA) models**

The second method of modeling which is used is the Auto-Regressive Moving-Average (ARMA) method. ARMA models are widely used in hydrology, dendrochronology, and many other fields [5]. Auto-regressive Moving-average (ARMA) models are mathematical models of the persistence, or autocorrelation, in a time series[6]. ARMA models can effectively be used to predict behavior of a time series from past values alone.
ARMA models are a generalization of Auto-Regressive (AR) and Moving Average (MA) models and a special case of Auto-regressive Integrated Moving Average (ARIMA) models.

**The Method Mathematically**

Auto-Regressive (AR) models and Moving-Average (MA) models are combined to form one model called Auto-Regressive Moving Average (ARMA) model. The order of AR model is the time steps which the model will go back to predict the future value and the order of MA model is the past difference steps which the model will go to predict the future value. The auto-regressive model includes lagged terms on the time series itself, and that the moving-average model includes lagged terms on the noise or residuals. Combing the lagged terms; it gives what are called auto-regressive moving-average, or ARMA, models. The order of the ARMA model is included in parentheses as ARMA (p, q), where p is the auto-regressive order and q the moving-average order. The simplest and most frequently used ARMA model is ARMA (1, 1) model [7].

$$\theta_t + a_1\theta_{t-1} = \epsilon_t + b_1\epsilon_{t-1} \quad (1)$$

Where $\theta_t$ is the mean-adjusted series in year t, $\theta_{t-1}$ is the series in the previous year i.e. lag of one, $a_1$ is the lag-1 autoregressive coefficient, Where $\epsilon_t$ and $\epsilon_{t-1}$ are the residuals or the noise or the random-shock at times t and t-1, and $b_1$ is the first-order moving average coefficient. The residuals $\epsilon_t$ are assumed to be random in time (not auto-correlated), and normally distributed. General form of ARMA (p, q) model is given below:

$$[1 - \sum_{s=1}^{P} \Psi_s(L)s] \theta_t = [1 - \sum_{s=1}^{q} \phi_s(L)s] \epsilon_t \quad (2)$$

p and q are called orders of the model, $\Psi_s$ is called AR model coefficients and $\phi_s$ is called MA model coefficients, L is an operator called the 'lag operator' when it operates on any data points; the result is the previous data value. Mathematically:

$$L^s(\theta_t) = \theta_{t-s} \quad (3)$$

In short form the ARMA (p, q) model is written as:

$$\Psi(L)\theta_t = \phi(L)\epsilon_t \quad (4)$$

This is most general form of ARMA (p, q)

Auto-Regressive Integrated Moving-Average (ARIMA) models the generalized forms of ARMA models. They are applied in some cases where data shows the evidence of non-stationary [18], where an initial differencing step can be applied to remove the non-stationary; the differencing step corresponds to 'integrated' part of ARIMA. The general form can be written as

$$\Psi(L)\{1 - L\}^d \theta_t = \phi(L)\epsilon_t \quad (5)$$

Where d is the level of differencing which determines the number of differences taken to remove the non-stationary. Stationary in simple terms implies that the probability density of the data does not change when data is shifted in time or space. The condition which is necessary is that the mean and variance (if they exist) should remain same when data is shifted.

**Order Selection**

There are different methods to determine the orders of ARMA model. Some of the are auto-correlation function(ACF), partial auto-correlation function (PACF).Order selection can be made on the basis on the model validity criteria; such as the Akaike's information criterion (AIC) and the minimum description length (MDL) [8].

**Case study**

The data and station is same as taken for the polynomial curve fitting model. The general equation for the ARMA (p, q) model is given below:

$$\Psi(L)\theta_t = \phi(L)\epsilon_t \qquad \text{Which is equation (1)}$$

For the present study ARMA (1, 1) and ARMA (2, 2) are used for simplicity; also because the error is not very sensitive to small changes in the order model around the order that provides minimum error [9].

**Case study model and results:
ARMA(1,1) and ARMA(2,2) are considered**

| MODEL | ψ(L) | Φ(L) |
|---|---|---|
| ARMA(1,1) | $1 - 0.9876 L^{-1}$ | $1 - 0.2108 L^{-1}$ |
| ARMA(2,2) | $1 - 1.552 L^{-1} + 0.5526 L^{-2}$ | $1 - 0.788 L^{-1} - 0.06111 L^{-2}$ |

| time ahead | Model | Mean square error (MSE) | Correlation coefficient (r) |
|---|---|---|---|
| 3 hour ahead | **ARMA(1,1)** | **0.0632** | **0.996** |
| | **ARMA(2,2)** | **1.0873** | **0.922** |
| 6 hour ahead | **ARMA(1,1)** | **0.9738** | **0.9645** |
| | **ARMA(2,2)** | **1.0152** | **0.9123** |
| 12 hour ahead | **ARMA(1,1)** | **0.9362** | **0.9090** |
| | **ARMA(2,2)** | **1.1524** | **0.88524** |

If the absolute values of each of operating functions {ψ(L), Φ(L) }have terms (ignoring the initial unity value) must be less than 1 then the data can be said to be stationary[10].

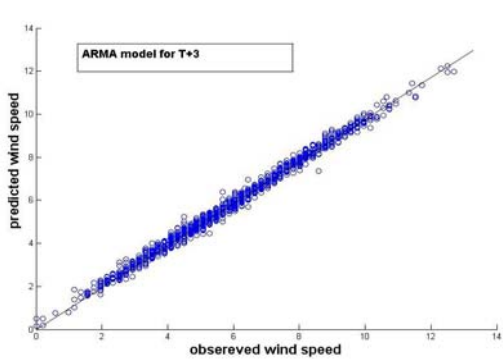
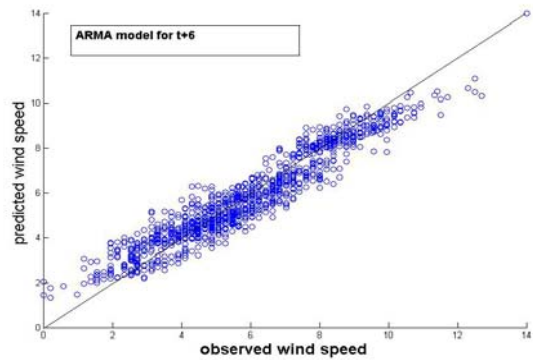

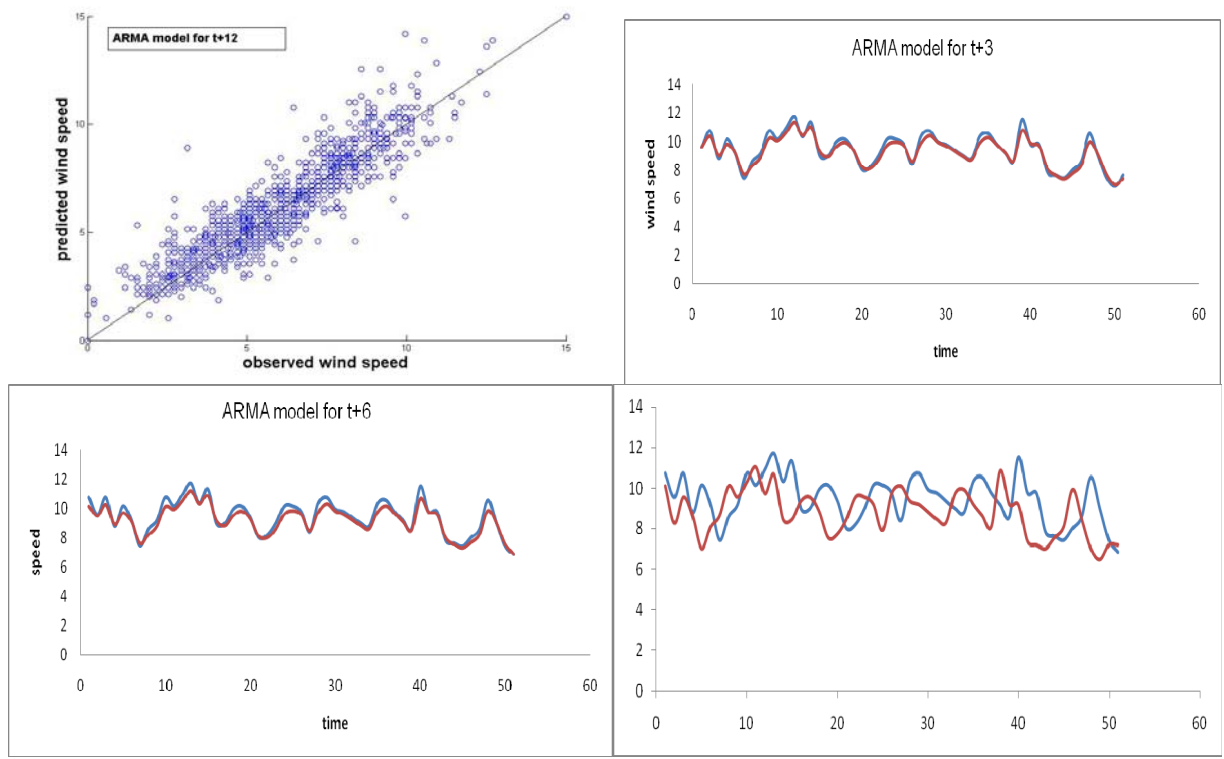

**Figure 2** Comparison of observed wind speed and predicted wind speed for 3, 6 and 12 hour ahead wind speed prediction with ARMA models from "01/08/2001 3:15" to "16/12/2001 12:15".

Noticing the above models and corresponding figures it is clear that the correlation is very high for prediction of t+3 wind speed where as for the prediction of wind speeds for longer intervals the correlation gets decreased and so is for the mean square error.

**The Artificial Neural Networks (ANN) model**

**Introduction**

Neural network models are the recently developed models and have gained a lot of fame in a little period of time. These models are used in a large number of applications. The engineering applications include function approximation, optimization, system modelling and pattern recognition[11]. Artificail nerural networks are used mostly for modelling in hydrlogy. Artificial neural networks have many advantages over the staistical, stochastic, and deterministic approaches[11].

Some of the advantages are:
1. The networks do not assume any mathematical mode a prori as is the case with the other approaches.The results depend on the data and can recognize the pattern without learning definitions.

2. Small input errors can be tollerated because the distributed processing ensures that significant change in the out put does not happen in that way[11].

3. Machine learning systems are also easier to reconfigure, new sets of examples can be quickly generated.

4. Ann's have the ability to cater to irregular seasonal variation in the data with the aid of their hidden layer nodes [12].

**Definition**

Artificail neural networks are similar to the biological neural networks that are present in brain.Artificial neural networks are the networks constructed by connecting artificial neurons. Artificial neural networks are input out networks[13].

**Neurons**

McCulloch and Pitts first introduced the concept of artificial neuron in 1943.However it was only after presentation of the network-training scheme of back propagation by Rumelhart et al (1986), that number of applications started.
Neurons are the main elements in an artificail neural network.These are compution and information processing units.The various elements of neural model are:
Connecting link or a set of synapses.
Summation point also called adder.
activation function.

*a. Connection links*

All the links are charectarised by weights (its strength).A signal to the link say $x_j$; x input of j link is connected to the neuron say k.The signal $x_j$ is multiplied by the link weight ($w_{kj}$); here first subscript k is the neuron to which link kj is connected and xj is the output to link j.

*b. Adder*

It is used for summing the input signal to neuron i.e weighted by respective links (synapses) of neuron.

*c. Activation function*

It is used to give the output from neuron.It can be sigmoidal fuction,sinusoidal function, gussian or hyperbolic function. The activation function gives output.
An activation function is for limiting the amplitude of output.The structure of neural model is shown below

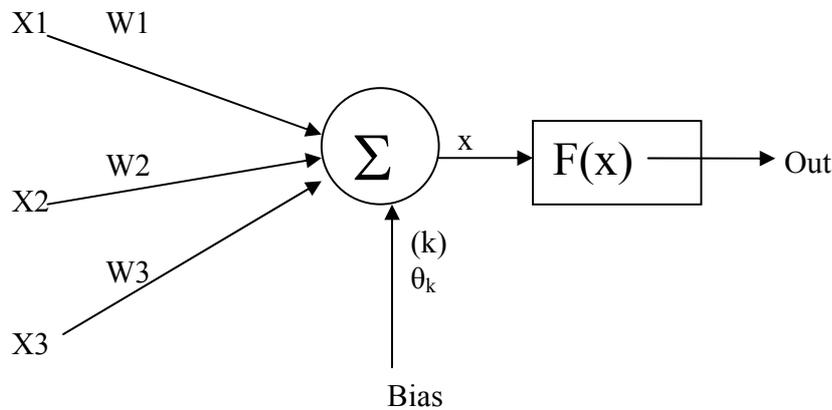

*Figure. 3. Structure of a neural model:*

$$\Sigma = \sum_{i=1}^{n}(w_{kj}x_j) \quad (1)$$
$$S = \Sigma + \theta_k \quad (2)$$
$$\theta = 1/(1 + e^{-s}) \quad (3)$$

Each neuron is represented by a single node called computational node.

**Networks**

**Feedforward Networks**

In this type of network the connections between neurons flow in one dfirrection i.e. from input layer through hidden layer(s) to output layer.Feedforward is said to exist in a dynamic behavior when the output of an element in a system influences inpart the input applied to that particular element hence giving rise to one or more closed paths for transmission of signals.It can be of sigle layer or multiple layer type.

  a. *Single layer feedforward*

In a single layer network there is only one output (computational)layer[14].Since in input layer there there is no computation it is not counted.

  b. *Multilayer feedforward networks*

There are one or more hidden layers present, whose computatinal nodes are called hidden neurons.By drawing more hidden layers network is enabled to extract higher order statistics.

**Recurrent networks**

Recurrent network is different from feedforward in a sense that it has at least one feedback loop i.e.the output is fed back to preceding layer or is fedback to the layer itself; which is called self feedback.The presence of feedback has a profound impact on learning capability of network.

**Nueral Netwok process**

The whole neural network process may be devided into branches,which are given as:

*Choice of performance criteria*

The performance critreria may be :
Prediction accuracy and/or Training speed and/or Delays which is dependent on processing speed.

*Devision of data*

The common practice is to split the data into two subsets:
a. A training set: traning set is used to train the network.

b. An independent validation set: validation set is used to validate the results/outputs.

And many a times third set called cross-validation set is also adopted; which is used for a cross-checking at various stages of training and learning.

*Data processing*

In order to ensure that all the variables receive equal attention during training they should be standardrized for example; the outputs of the logistic transfer function are between 0 and 1; the data is generally scaled in the range 0.1 to 0.9 or .2 to .8[12]. Removal of deterministic components should be considered.Various techniques like classical decomposition, adaptive weight updata strategy may be employed.However this can be also be omitted.

*Determination of network archtecture*

Network architecture determines the number of connection weights and the way information is processed through the network.Determination of an appropraite network architecture is one of the most important and difficult task in the model building process. It consists of:

a. *Type of connection and degree of connectivity*

Feedforward networks are mostly used; in which one layer nodes are connected to the nodes in next layer.Recurrent networks are recently been used.Here nodes on one layer can be connected to the nodes in the next layer, previous layer and even to themselves.
The degree of connectivity (i.e. whether the nodes in a network are fully or partailly connected) has an impact on number of connection weights that need to be optimized[11]. In most of the cases fully connected networks are employed.

b. *Geometry*

Network geometry provides the number of connection weights and how these are arranged.This is generally done by fixing the number of hidden layers and choosing the number of nodes in each layer..Traditionally optimal network geometrics have beem found by trail and error method.There are some systematic methods also[11].
   a. Pruning algorithms
   b. Constructive algorithms

*Optimization or training or learning*

The processe of optimizing the connection weights is known as training or learning. Non-linear optimization techniques are used. Optimization can be carried out using local or global methods. Global methods have the ability to escape local minima in the error surface and are thus able to find the optimal or near optimal weight configurations[15].

Stopping criteria: The criteria used to stop the traning process are virtually important because dependig on this, model can be optimally or sub-optimaly trained the training should neither stop too early nor once overfitting of data has occurred.

*Validation*

After the training phase is complete the performance of the model has to be checked out or validated.This is done using independent data set. Error function used is the Mean square error (MSE).There are a other error measures also which are given below[16]:
1. Mean squared relative error (MSRE).
2. Coefficient of efficiency (CE).
3. Cofficient of determinatiom ($r^2$).

**Case study results**

Numerious trails are done to get the best network model. The trails with corresponding correlation coefficients and mean square error(MSE) are given as under:
Here only the best ones are shown: Feedforward Backpropagation network was choosen for the study.Number of input nodes are fixed, what can be varied is the neurons in hidden layers[17].

**Case study models and results:**

| Time ahead | No. of input nodes | No. of hidden layers | No. of neurons in hidden layer | Transfer function in input layer | Transfer function in hidden layer(s) | Training function | Mean square error(MSE) | Correlation coefficient (r) |
|---|---|---|---|---|---|---|---|---|
| 3 hour | 2 | 1 | 3 | purelin | logsig | trainlm | 0.2994 | 0.98945 |
| 6 hour | 2 | 1 | 5 | tansig | purelm | trainlm | 0.7936 | 0.9796 |

| 12 hour | 2 | 2 | 3(1$^{st}$) 1(2$^{nd}$) | logsig | Logsig (both) | trainscg | 1.1900 | 0.96065 |

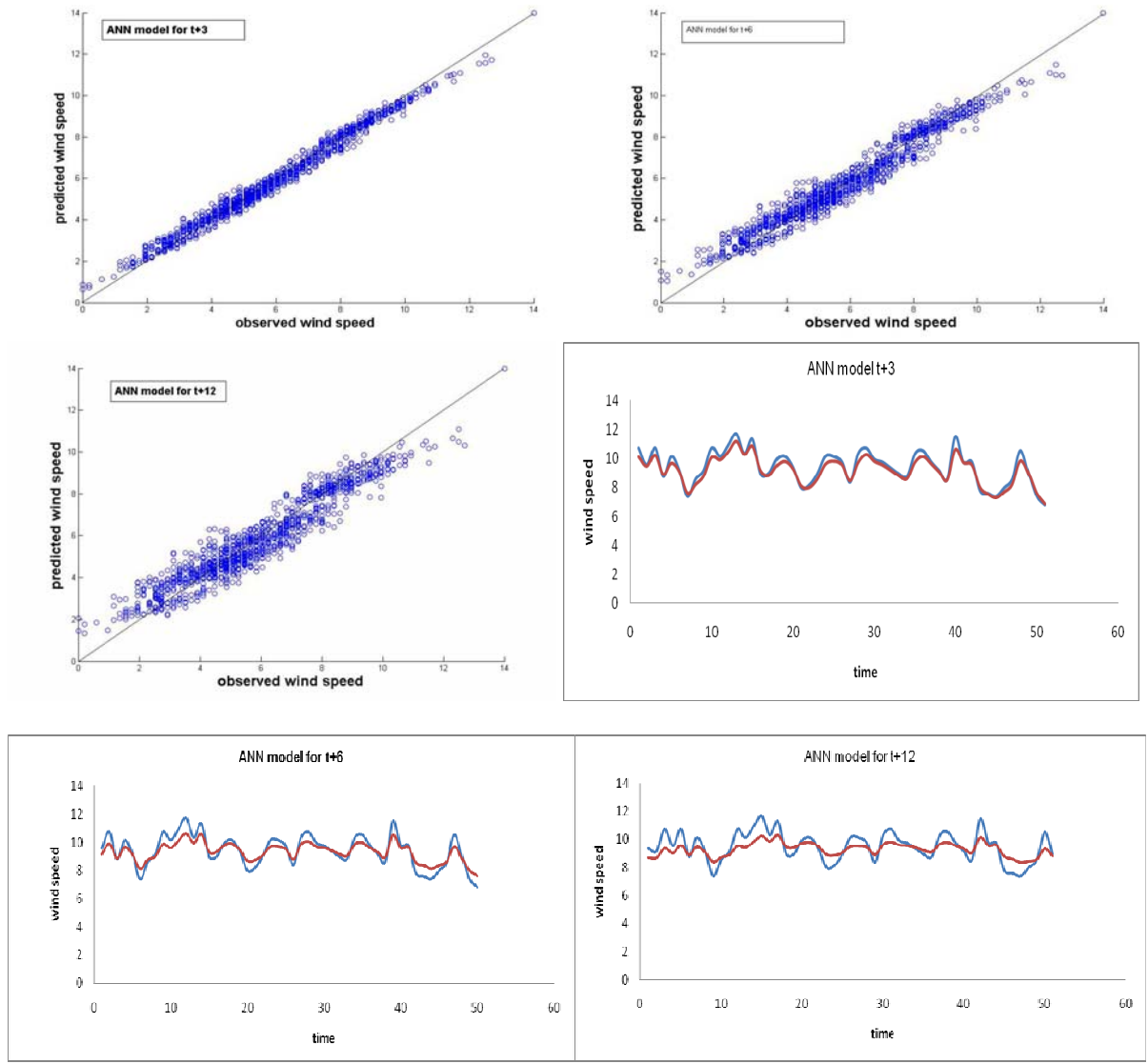

**Figure 4** Comparison of observed wind speed and predicted wind speed for 3, 6 and 12 hour ahead wind speed prediction with Artificial Neural Networks models from "01/08/2001 3:15" to "16/12/2001 12:15".

### Conclusion

Accurate estimation of wind speed distribution is critical to the assessment of wind energy potential, the site selection of wind farms, and the operations management of wind power conversion [18] and for onshore and offshore works[19].Polynomail curve fitting models were not foundto be so so accurate and should be avoided for use. ARMA models can be used for short term wind prediction and should not be used for long term wind prediction these models however can be improved significantly by using Bayesian approach[20]. ARMA models have an advantage that they can provide very accurate results for short term wind prediction if data have fairly smooth trend and stationarity. If the data are not stationarity and there are quite high varaitions and change trend suddenly they may lead to very inaccurate results. All the disadvantages above are fairly solved by using Artificial Neural Networks. Artificial Neural Networks depend on the training data and not on any physical relation.They take less amount of time in learning and giving results. The best network can be choosen by trial and error method. The results showed that neural networks can gave most accurate results for the long term wind speed prediction.The modelling of wind speed can be improved by using kalman filters or Baysien methodology or by using hybrid methods.There is a great need of improving and developing new modelling techniques for the sake of society and for preventing losses which take place due to imperfect prediction of wind speed. For long term wind prediction bayesian model averanging can be employed. Threre is a lot of scope in wind prediction and what is needed is the accuracy of forecast.